\documentclass[reprint,amsmath,amssymb,aps,prl]{revtex4-2}
\usepackage{graphicx,color}% Include figure files
\usepackage{dcolumn}% Align table columns on decimal point
\usepackage{bm}% bold math
\usepackage{easyReview} % provides commands \highlight, \comment, etc.
\begin{document}

%\linenumbers

\title{Observation of an electric quadrupole transition in a negative ion:\\  Experiment and Theory}% Force line breaks with \\
\author{C. W. Walter, S. E. Spielman, R. Ponce, N. D. Gibson}
\affiliation{Department of Physics and Astronomy, Denison University, Granville, OH  43023, USA}
\author{J. N. Yukich}
\affiliation{Physics Department, Davidson College, Davidson, North Carolina 28035, USA}
\author{C. Cheung$^1$ and M. S. Safronova$^{1,2}$}
\affiliation{$^1$Department of Physics and Astronomy, University of Delaware, Newark, Delaware 19716, USA\\
$^2$Joint Quantum Institute, National Institute of Standards and Technology and the University of Maryland, College Park, Maryland 20742, USA
}

\date{\today}% It is always \today, today,

\begin{abstract}
The first direct experimental observation of an electric quadrupole (\textit{E}2) transition between bound states of an atomic negative ion has been made.  The transition was observed in the negative ion of bismuth by resonant (1+1) photodetachment from Bi$^-$ $^3$\textit{P}$_2$ via excitation of the Bi$^-$ $^3$\textit{P}$_0$ fine structure state.  The \textit{E}2 transition properties were independently calculated using a hybrid theoretical approach to account for the strong multi-level electron interactions and relativistic effects. The experimental and theoretical results are in excellent agreement, providing valuable new insights into this complex system and forbidden transitions in negative ions.
\end{abstract}

\maketitle

Although optical transitions between bound states of both neutral atoms and positive ions have been known and extensively studied for more than a century, similar transitions between bound states of \textit{negative} ions have only been observed much more recently \cite{Andersen2004}.  This difference in discovery date is due, in part, to the fundamental nature of negative ions:  In sharp contrast to the infinite number of bound states in atoms and positive ions, since negative ions are not held together by a net Coulomb potential, their short-range polarization potentials can support only one or at most a few bound states \cite{Andersen1999,Pegg2004} making possible transitions very scarce.

The first observation of a bound-bound electric dipole (\textit{E}1) transition in an atomic negative ion was reported only 20 years ago (in Os$^-$) \cite{Bilodeau2000}, and magnetic dipole (\textit{M}1) transitions in negative ions in the optical regime had been first observed just 4 years earlier (in Ir$^-$ and Pt$^-$) \cite{Thogersen1996,Mader1972}.  Since these initial discoveries, \textit{E}1 transitions have been observed in two other negative ions (Ce$^-$ \cite{Walter2011} and La$^-$ \cite{Walter2014}), and \textit{M}1 transitions have been observed in several additional negative ions \cite{Scheer1997,Scheer1998}.  However, previous experimental searches for electric quadrupole (\textit{E}2) transitions in negative ions have not been successful, due to the very small transition rates and large background signals from continuum photodetachment \cite{Thogersen1996,Scheer1998}.  In the present work, we report the first direct experimental observation of an \textit{E}2 absorption transition between bound states of an atomic negative ion (Bi$^-$), together with high-precision theoretical calculations of the energy and rate of the observed transition.

Electric quadrupole atomic transitions are of great interest due to applications such as tests of fundamental physics \cite{Berengut2010,Dzuba2015,sanner2019}, optical clocks \cite{Ludlow2015}, and quantum information \cite{Schindler2013}, and they provide important benchmarks for detailed state-of-the-art theoretical calculations \cite{Safronova2017}.  The properties of negative ions crucially depend on electron correlation effects \cite{Fischer1989,Andersen2004,Ivanov2004,Pegg2004,Eiles2018}, and \textit{E}2 transitions in negative ions provide uniquely valuable opportunities to gain insights into these subtle but important interactions.  Accurate theoretical computations are very difficult for negative ions with complex electronic structure due to large configuration mixing in comparison with neutrals or positive ions \cite{Fischer1989,Andersen2004}.

There is even greater urgency for studying forbidden transitions in negative ions with the advent of new cryogenic storage ring facilities, such as DESIREE \cite{Thomas2011,Backstrom2015} and the CSR \cite{vonHahn2016,Meyer2017a}, that can measure lifetimes of excited states of negative ions over unprecedentedly long scales of up to hours \cite{Backstrom2015}.  While most of the negative ion excited state lifetime experiments to date have involved \textit{M}1 transitions, one recent study at DESIREE measured the \textit{E}2 decay of an excited state of Pt$^-$ \cite{Chartkunchand2016}.

The present work investigates an \textit{E}2 transition in the negative ion of bismuth both experimentally and theoretically.  The hyperfine-averaged binding energy of the Bi$^-$ (6\textit{p}$^4$ $^3$\textit{P}$_2$) ground state relative to the Bi (6\textit{p}$^3$ $^4$\textit{S}$_{3/2}$) ground state was previously measured by Bilodeau and Haugen to be 942.369(13) meV \cite{Bilodeau2001}.  While there have not been any previous measurements of the fine structure of Bi$^-$, Su \textit{et al.} very recently reported calculations that indicated an interesting inversion in the ordering of the excited fine structure levels, with $^3$\textit{P}$_0$ being bound and $^3$\textit{P}$_1$ unbound \cite{Su2019}.

We have carried out two complementary experiments in the present work (see Fig. 1):  (1) Measurement of the binding energy of the Bi$^-$ $^3$\textit{P}$_0$ excited state using photodetachment threshold spectroscopy; and (2) Observation of the Bi$^-$ $^3$\textit{P}$_2$ $\rightarrow$ $^3$\textit{P}$_0$ transition via resonant (1+1) photon detachment.  The observed resonance transition has
$\lvert\Delta\textit{J}\rvert$=2; therefore, it is strictly forbidden for \textit{E}1 or \textit{M}1 processes, and it must proceed by the \textit{E}2 interaction. Although there have been previous observations of transitions in negative ions that had both \textit{M}1 and \textit{E}2 contributions \cite{Scheer1997,Scheer1998}, to our knowledge, this is the first transition observed in a negative ion with \textit{E}2 as the lowest-order allowed interaction.  The Bi$^-$ fine structure and \textit{E}2 transition properties were independently calculated using a high-precision hybrid theoretical approach to account for the strong multi-level electron interactions and relativistic effects.  The experimental and theoretical results are in excellent agreement, providing valuable new insights into this complex system and testing the accuracy of the theoretical approach.

\begin{figure} [bt]
\includegraphics[width=86mm]{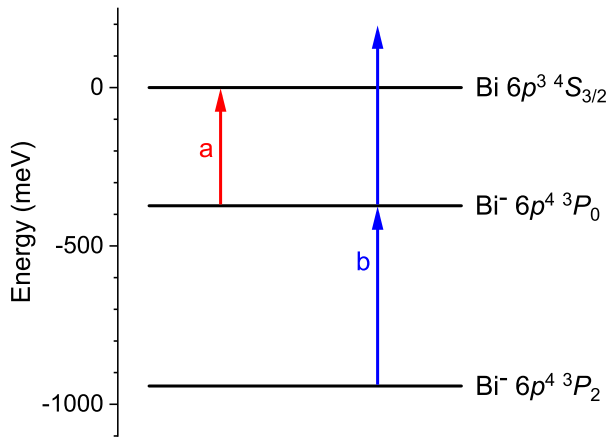}
\caption[]{\label{Fig_1}Energy levels of Bi$^-$ and Bi with arrows showing the two measurements performed: (a, red) Single-photon threshold detachment from Bi$^-$ $^3$\textit{P}$_0$; and (b, blue) Resonant (1+1) photon  detachment via the Bi$^-$ $^3$\textit{P}$_2$ $\rightarrow$ $^3$\textit{P}$_0$ \textit{E}2 transition.}
\end{figure}

Photodetachment from Bi$^-$ was measured as a function of photon energy using a crossed ion-beam—laser-beam system described previously \cite{Walter2010,Walter2011}.  Negative ions produced by a cesium sputter source \cite{Middleton1990} using a cathode packed with bismuth powder were accelerated to 12 keV and magnetically mass selected producing $\sim$1 nA of $^{209}$Bi$^-$ (the only long-lived naturally-occurring isotope).  The ion beam was intersected perpendicularly by a pulsed laser beam, following which residual negative ions were electrostatically deflected into a Faraday cup, while neutral atoms produced by photodetachment continued undeflected to a multi-dynode detector.  The neutral atom signal was normalized to the ion-beam current and the photon flux measured for each laser pulse.  The spectra were obtained by repeatedly scanning the laser wavelength over a range and then sorting the data into photon energy bins of selectable width.

The laser system consisted of an optical parametric oscillator-amplifier (OPO-OPA) (LaserVision) pumped by a 20-Hz Nd:YAG laser, giving an operating range of 5000 – 1300 nm (250 – 920 meV).  Broad survey scans were performed with the pump laser operating broadband giving a bandwidth of $\sim$0.1 meV, while narrow scans near sharp structures were performed with injection seeding of the pump laser to reduce the bandwidth to $\sim$0.01 meV.  The laser beam diverges slightly as it leaves the OPA, so long focal length lenses were used to approximately collimate the beam.  For the \textit{E}2 transition measurement, an additional 50-cm focal length lens was placed about 40 cm from the interaction region to increase the intensity to drive this very weak transition. 

\begin{figure} [bt]
\includegraphics[width=86mm]{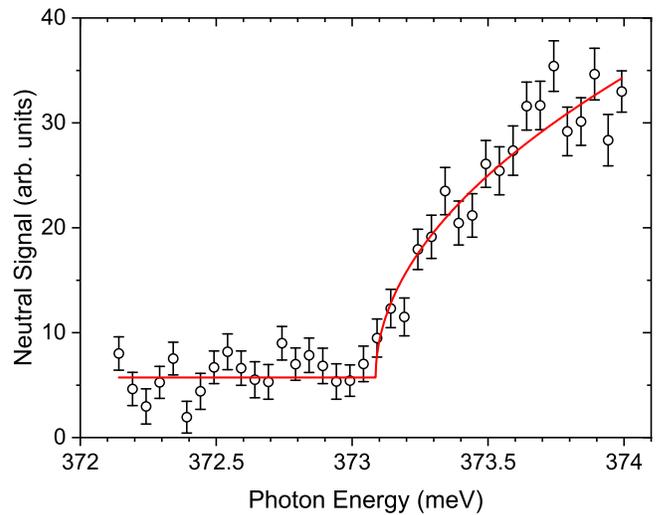}
\caption[]{\label{fig2}Measured photodetachment threshold from the Bi$^-$ $^3$\textit{P}$_0$ excited state to the Bi (6\textit{p}$^3$ $^4$\textit{S}$_{3/2}$) ground state; (circles) data , (line) \textit{s}-wave Wigner law fit.}
\end{figure}

The photodetachment spectrum from Bi$^-$ was measured from 255 meV to 920 meV, revealing only a single threshold near 373 meV, which is due to detachment from the Bi$^-$ (6\textit{p}$^4$ $^3$\textit{P}$_0$) fine structure excited state to the Bi (6\textit{p}$^3$ $^4$\textit{S}$_{3/2}$) ground state.  Fig. \ref{fig2} shows the measured neutral atom signal in the vicinity of the observed threshold.  It is worthwhile to note that most of the ions in the beam are in the $^3$\textit{P}$_2$ ground state rather than the $^3$\textit{P}$_0$ excited state.  An estimate based on the relative photodetachment signal rates 0.5 meV above threshold in the present experiment compared to our recent measurements of Tl$^-$ \cite{Walter2020} indicates that Bi$^-$ $^3$\textit{P}$_0$ makes up only $\sim$0.3\% of the beam, which is also consistent with estimates based on a Boltzmann statistical distribution at the approximate ion source temperature of $\sim$1500 K.  Also, note that there is a small background signal below the $^3$\textit{P}$_0$ threshold; since no more weakly bound excited states of Bi$^-$ are expected, this background is most likely due to a slight leakage of the OPO ``signal'' light at a photon energy of 1.96 eV which is sufficient to photodetach the large population of ground state Bi$^-$ ions in the beam.

The threshold energy for Bi$^-$ (6\textit{p}$^4$ $^3$\textit{P}$_0$) detachment can be precisely determined from the data in Fig. 2 using the Wigner threshold law \cite{Wigner1948}.  In the present case, a \textit{p} electron is detached, so the cross section closely above threshold is dominated by \textit{s}-wave detachment and increases as (\textit{E} - \textit{E$_t$})$^{1/2}$, where \textit{E} is the photon energy and \textit{E$_t$} is the threshold energy .  The Wigner law provides an excellent fit to the data, yielding the detachment threshold corresponding to the binding energy of Bi$^-$ $^3$\textit{P}$_0$ to be 373.09(4) meV. 

\begin{figure} [bt]
\includegraphics[width=86mm]{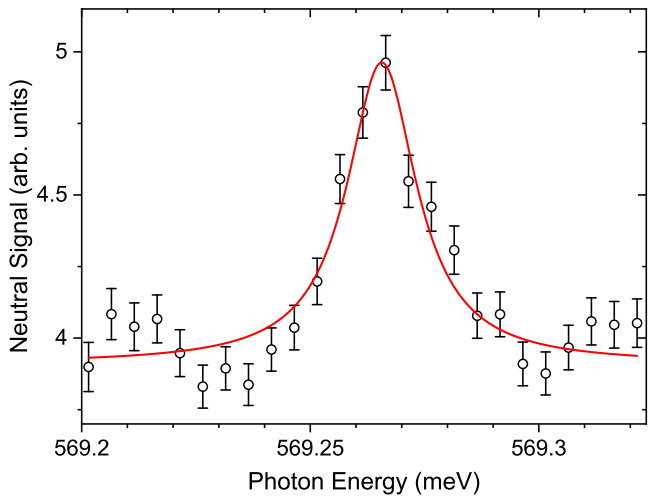}
\caption[]{\label{fig3}Measured peak for resonant (1+1) photodetachment via the Bi$^-$ $^3$\textit{P}$_2$ $\rightarrow$ $^3$\textit{P}$_0$ \textit{E}2 transition.   The solid line is a Lorentzian fit to the data.}
\end{figure}

After establishing the binding energy of the excited state, we were able to search for and find the Bi$^-$ $^3$\textit{P}$_2$ $\rightarrow$ $^3$\textit{P}$_0$ electric quadrupole transition.  The expected transition energy of 569.28(4) meV is given by the difference between the binding energies of the Bi$^-$ $^3$\textit{P}$_2$ ground state (hyperfine-averaged value 942.369(12) meV \cite{Bilodeau2001})  and the Bi$^-$ $^3$\textit{P}$_0$ excited state (373.09(4) meV).  Figure \ref{fig3} shows the neutral atom signal measured in the vicinity of the expected transition; a peak in the signal is visible due to resonant (1+1) photodetachment from Bi$^-$ $^3$\textit{P}$_2$ via \textit{E}2 excitation of Bi$^-$ $^3$\textit{P}$_0$ followed by absorption of a second photon to detach the excited state.  Observation of this resonance transition in the present experiment was possible in part because the single-photon detachment background signal is low due to the small fraction of excited state ions in the beam.  A Lorentzian fit to the measured data yields the transition energy to be 569.27(3) meV, in excellent agreement with the expected value of 569.28(4) meV based on the difference in binding energies.  The fitted peak width of 0.023(5) meV is an upper limit rather than the natural width of the transition due to the lifetime of the upper state, because of broadening by the bandwidth of the OPA ($\sim$0.01 meV) and by unresolved hyperfine structure.

The measurements reported in this work provide an excellent opportunity to test the accuracy of theory in this complicated negative ion.
We carried out calculations of the Bi$^-$ binding energies and the $^3$\textit{P}$_2$ $\rightarrow$ $^3$\textit{P}$_0$ transition energy and rate using a hybrid approach that combines the configuration interaction (CI) and the coupled-cluster method (CI+all-order method) \cite{SafKozJia09}. This method allows efficient inclusion of both  strong valence-valence correlations (via the CI)
and core excitations from the entire core (by the coupled-cluster approach).
The many-electron wave function is obtained as usual in the framework of the CI method as a
linear combination of all distinct many-electron states of a given angular momentum $J$ and parity:
$ \Psi_{J} = \sum_{i} c_{i} \Phi_i$
, but the energies and wave functions of the low-lying states are
determined by diagonalizing the effective (rather than bare CI) Hamiltonian. It is constructed using a combination of the coupled-cluster approach that allows single and double excitation from the entire core and the many-body perturbation theory (MBPT).
Alternatively, we carried out identical computations constructing the entire effective Hamiltonian using the second-order MBPT \cite{KozPorSaf15} to evaluate the importance of the higher-order corrections; we refer to such results as the CI+MBPT.

We treat Bi as a  system with 3 valence electrons and a [Xe]$4f^{14}5d^{10}6s^{2}$ core. The core is the same for the Bi$^-$ calculation. The difference between the Bi$^-$ and Bi calculation is in the CI part, which contains 4 valence electrons for Bi$^-$. There is an exponential growth in the number of possible configurations with the addition of extra valence electrons and care must be taken to ensure a sufficiently large set of configurations for Bi$^-$. The problem is exacerbated for the weakly-bound negative ion which exhibits very strong configuration mixing.
All calculations incorporate the Breit interaction as described in \cite{PorSafSaf20}. 

\begin{table*} [bt]
\caption{\label{table1} Calculated binding energies of Bi$^-$ bound states relative to the Bi $^4$\textit{S}$_{3/2}$ ground state and transition energy for Bi$^-$ $^3$\textit{P}$_2$ $\rightarrow$ $^3$\textit{P}$_0$ compared to experimental measurements in meV. Contributions of the higher orders (HO) are calculated as the differences of the CI+all-order (CI+all) and the CI+MBPT calculations.
QED corrections and contributions of the higher partial wave (l=6) are listed separately. The column ``Extra conf.'' lists the difference of the results of the large and medium CI calculations.
``Final'' results were obtained by including all of these corrections in the same computation. ``Diff.'' and ``Diff. \%'' are the difference between the experimental and final calculated energies. See text for more explanations.
}
\begin{ruledtabular}
\begin{tabular}{llccccccccccc}
\multicolumn{2}{c}{}&\multicolumn{1}{c}{$J$}& \multicolumn{1}{c}{Expt.}& \multicolumn{1}{c}{CI+MBPT}&  \multicolumn{1}{c}{CI+all}&
\multicolumn{1}{c}{HO}& \multicolumn{1}{c}{$l=6$}& \multicolumn{1}{c}{QED}& \multicolumn{1}{c}{Extra conf.}& \multicolumn{1}{c}{Final}& \multicolumn{1}{c}{Diff.}&  \multicolumn{1}{c}{Diff.\%} \\
\hline
Bi$^-$  & $6p^4$ $^3$\textit{P} &  2&  942.369(13)$^a$  & 910.4  & 911.0  & 0.6  & 2.1  & 2.1  & 26.7  & 941.9  & 0.5  &  0.05\%  \\
& $6p^4$ $^3$\textit{P} &  0 & 373.09(4)$^b$  &  321.2  &  342.6  &  21.3  &  1.0  &  1.4  &  26.0  &  371.0  &  2.1  &  0.6\%  \\[0.5pc]
&$^3$\textit{P}$_2$ $\rightarrow$ $^3$\textit{P}$_0$ &    &    569.27(3)$^b$  & 589.2  & 568.5  & -20.7  & 1.1  & 0.7  & 0.6  & 570.9  & -1.7  &  0.3\%

\end{tabular}
\end{ruledtabular}
\flushleft $^a$ Ref.~\cite{Bilodeau2001}, hyperfine-averaged.\\
$^b$ Present measurements.
\end{table*}

The results of the CI+all-order and CI+MBPT calculations and specific contributions to the energies are summarized for Bi$^-$ in Table~\ref{table1}; similar results for neutral Bi are given in the Supplemental Materials \cite{Supplemental2020}.  Binding energies are shown relative to the Bi $6p^3$ $^4$\textit{S}$_{3/2}$ ground state.  Contributions of the higher orders (HO) are calculated as the differences of the CI+all-order and the CI+MBPT calculations.  To evaluate the accuracy of the calculations, we calculated several smaller corrections separately. We originally ran CI+all-order and CI+MBPT calculations allowing excitations to all partial waves up to $l=5$, with maximum principal quantum number $n=35$ for each (relativistic) partial wave. The contribution of the $l=6$ partial wave is listed in the column ``$l=6$''. From the extrapolations carried out for simpler systems, we find that the contribution of all other partial waves is on the same order as the $l=6$ contribution. QED corrections were calculated following \cite{TupKozSaf16}. Both of these corrections are relatively small. Next, we increase the number of CI configurations allowing excitations up to $23spdf18g$ and  $22spdf18g$ orbitals for Bi and Bi$^-$, respectively, an increase from the initial  $22spd18f14g$ set. All single, all double, and a large subset of triple excitations are included.  These changes increase the number of included configurations for Bi$^-$ from 73\:719 to 126\:168, with corresponding increase in the number of Slater determinants from
3\:090\:923 to 4\:952\:692. Finally, we carried out a complete CI+all-order run that incorporated all corrections (QED, $l=6$, and larger number of configurations) simultaneously. These results are listed as ``Final'' in Table~\ref{table1}.

Our final calculated binding energy of the Bi$^-$ $^3$\textit{P}$_2$ ground state is in excellent agreement with the measured value of Bilodeau and Haugen \cite{Bilodeau2001}, differing by only 0.5 meV or 0.05\% (see Table I).  We find that the binding energy of the ground state is strongly affected by the inclusion of more configurations but not by the inclusion of the higher orders. This is expected since the Bi and Bi$^-$ computations share the same core; thus, difference in its treatment is expected to cancel to a degree. A sensitivity to extra configurations is also expected  as configuration mixing for Bi$^-$ is much stronger than for Bi. There is also excellent agreement for the binding energy of the fine structure excited state $^3$\textit{P}$_0$, with our calculated value of 371.0 meV being within 2.1 meV or 0.6\% of our measured value of 373.09(4) meV.  However, in contrast to the ground state, higher orders contribute significantly (5.7\%) to the binding energy of the $^3$\textit{P}$_0$ state and therefore affect the $^3$\textit{P}$_2$ $\rightarrow$ $^3$\textit{P}$_0$ transition energy.  Finally, our calculations indicate that the Bi$^-$ $^3$\textit{P}$_1$ state is not bound, which is in agreement with the calculations of Su \textit{et al.} \cite{Su2019}.  

It is interesting to also explore if Po, which has the same two lowest electronic states as Bi$^-$, may be used as a homologue system to improve prediction for the negative ion. We carried out a Po computation with all parameters identical to Bi$^-$; the detailed results are listed in the Supplemental Materials \cite{Supplemental2020}.  We find that the difference with experiment is actually larger in Po than in Bi$^-$. Most likely, this is due to uneven cancellation of some omitted effects, such as core triple excitations and non-linear terms that tend to strongly cancel.
We also find as expected that there is much stronger configuration mixing in the negative ion; for example, only 11 non-relativistic configurations  contribute a total of 99\% for the ground state of Po, but 22 configurations are needed for Bi$^-$. Our results demonstrate the significant fact that an isoelectronic neutral system cannot always be used as a homologue for a negative ion.

The present experimental and theoretical results for the Bi$^-$ $^3$\textit{P}$_2$ $\rightarrow$ $^3$\textit{P}$_0$ \textit{E}2 transition are given in Table~\ref{table3}, together with previous calculations.  Our calculated transition energy (570.9 meV) is in excellent agreement with our experimental value (569.27(3) meV), differing by only 1.7 meV or 0.3\%.  In contrast, the calculated transition energy from Su \textit{et al.} \cite{Su2019} is 55 meV larger than our measurement, while the earlier calculation of Konan \textit{et al.} \cite{Konan2013} is even farther away.
Our computations include higher-order inner-shell electronic correlations and, therefore,  are expected to be more accurate than  multiconfiguration Dirac–Hartree–Fock calculations \cite{Su2019,Konan2013} for both energies and transition rates.
\begin{table} [bt]
\caption{\label{table3} Present results and previous calculations for the Bi$^-$ $^3$\textit{P}$_2$ $\rightarrow$ $^3$\textit{P}$_0$ \textit{E}2 transition energy and upper-state lifetime.}
\begin{ruledtabular}
\begin{tabular}{lccc}
\multicolumn{1}{l}{Study}&\multicolumn{1}{c}{Method}& \multicolumn{1}{c}{$^3$\textit{P}$_2$-$^3$\textit{P}$_0$ Energy (meV)}& \multicolumn{1}{c}{Lifetime (s)}\\
\hline
Present&  Experiment &  569.27(3) &     -      \\
Present&  Theory&  570.9 & 16.5(7)                \\
Su \cite{Su2019} &  Theory & 624.0 & 15.20 \\    
Konan \cite{Konan2013} &  Theory & 1100.0 &   -  \\
\end{tabular}
\end{ruledtabular}
\end{table}

Turning now to the transition rate, we calculate the electric quadrupole $6p^4$~$^3\textit{P}_2-^3\textit{P}_0$ reduced matrix element to be 16.30(33)$ea_0^2$ using the CI+all-order method. There is only a 1\% difference between the CI+all-order and the CI+MBPT results, and there is a 1.7\% difference between the results obtained with medium and large sets of CI configurations. We add these in quadrature to estimate the final uncertainty of the matrix element to be 2\%. Using the experimental value of the transition energy, we obtain 0.0607(24)s$^{-1}$ for the transition rate, corresponding to a $^3$\textit{P}$_0$ lifetime of 16.5(7)~s.  At first glance, the previous calculated lifetime by Su \textit{et al.} \cite{Su2019} of 15.20 s appears to be fairly close to our value.  However, it is important to consider that Su \textit{et al.}'s quoted lifetime was obtained using their calculated transition energy for $^3$\textit{P}$_2$ $\rightarrow$ $^3$\textit{P}$_0$, which is larger than our precisely measured energy by $\sim$10\%.  Since the \textit{E}2 lifetime scales inversely with transition energy to the fifth power, revising their lifetime using our measured energy would yield an adjusted lifetime of 24.05 s, which is significantly longer than our value. Finally, note that although our calculated lifetime of the $^3$\textit{P}$_0$ upper state of 16.5(7) s indicates that the transition is far too narrow ($\sim$2x10$^{-14}$ meV) for a direct measurement of the lifetime from the peak width in the present experiment, the theoretically predicted lifetime could be rigorously tested in storage ring experiments using established techniques \cite{Backstrom2015, Chartkunchand2016}.  

In summary, we have measured the binding energy of the Bi$^-$ $^3$\textit{P}$_0$ excited state and observed its excitation from the Bi$^-$ $^3$\textit{P}$_2$ ground state using resonant (1+1) photodetachment.  To our knowledge, this is the first direct observation of an \textit{E}2 absorption transition between bound states of an atomic negative ion.  Furthermore, we have confirmed the fine structure of Bi$^-$ and the \textit{E}2 character of the transition through detailed theoretical calculations including the transition rate.  The measured and calculated energies are in excellent agreement, demonstrating the power of the theoretical methods used to account for the important correlation and relativistic effects in this complex multielectron system.      

Similar experimental and theoretical methods can be applied to study \textit{E}2 transitions in other negative ions that have appropriate excited bound state structures, opening a new avenue for investigations of forbidden transitions in atomic systems.  Such studies can be combined with the new ability to accurately measure lifetimes of excited negative ions over long time scales recently developed at cryogenic storage ring facilities such as DESIREE and the CSR to give further insights into many-body correlation effects and decay dynamics.
 
\begin{acknowledgments}
We thank Charles W.  Clark for insightful discussions.
This work was supported in part by U.S. NSF Grants No.\ PHY-1620687 and No.\ PHY-1707743. This research was performed in part under the sponsorship of the U.S. Department of Commerce, National Institute
of Standards and Technology. 
The theoretical research was supported in part
through the use of Information Technologies resources at
the University of Delaware, specifically the high-performance Caviness computing cluster.
\end{acknowledgments}

 \begin{table*} [b]
\caption{Supplemental Material. Calculated energy levels of Bi relative to the Bi $^4$\textit{S}$_{3/2}$ ground state in meV.  The experimental values are from NIST$^1$.  Contributions of the higher orders (HO) are calculated as the differences of the CI+all-order (CI+all) and the CI+MBPT calculations.  QED corrections and contributions of the higher partial wave (l=6) are listed separately. The column ``Extra conf.'' lists the difference of the results of the large and medium CI calculations.
``Final'' results were obtained by including all of these corrections in the same computation. ``Diff.'' and ``Diff. \%'' are the difference between the experimental and final calculated energies. See main text for more explanations.
}
\begin{ruledtabular}
\begin{tabular}{llccccccccccc}
\multicolumn{2}{c}{}&\multicolumn{1}{c}{$J$}& \multicolumn{1}{c}{Expt.}& \multicolumn{1}{c}{CI+MBPT}&  \multicolumn{1}{c}{CI+all}&
\multicolumn{1}{c}{HO}& \multicolumn{1}{c}{$l=6$}& \multicolumn{1}{c}{QED}&  \multicolumn{1}{c}{Extra conf.}& \multicolumn{1}{c}{Final}& \multicolumn{1}{c}{Diff.}&  \multicolumn{1}{c}{Diff.\%} \\
\hline
Bi     &  $6p^3$ $^4$\textit{S}       & 3/2&       0  &      0  &      0  &    0  &   0  &  0   &    0  &      0  &  0    &    -    \\
       &  $6p^3$ $^2$\textit{D}       & 3/2&  1415.8  & 1386.5  & 1386.1  & 0.4  & 3.5  & 1.1  & -4.2  & 1386.6  & 29.1  &  2.1\% \\
       &  $6p^3$ $^2$\textit{D}       & 5/2&   1914.0  & 1869.6  & 1874.3  & -4.7  & 4.3  & 1.0  & -10.3  & 1869.6  & 44.4  &  2.3\% \\
       &  $6p^3$ $^2$\textit{P}       & 1/2&   2685.6  & 2759.8  & 2699.0  & 60.8  & 4.8  & 1.5  & -13.0  & 2692.2  & -6.6  &  0.2\% \\
       &  $6p^3$ $^2$\textit{P}       & 3/2&   4111.9  & 4132.6  & 4041.4  & 91.3  & 9.2  & 3.0  & -10.7  & 4042.0  & 69.9  &  1.7\% \\
       &  $6p^2 7p$    & 1/2&   5098.8  & 5205.2  & 5114.1  & 91.1  & 5.6  & 2.5  & 38.1  & 5160.6  & -61.7 &  1.2\%  \\
       &  $6p^2 7p$    & 3/2&   5323.9  & 5421.8  & 5332.6  & 89.3  & 5.7  & 2.5  & 44.6  & 5385.7  & -61.8 &  1.2\%  \\
       &  $6p^2 8p$    & 1/2&   6198.9  & 6341.7  & 6253.4  & 88.3  & 6.3  & 2.5  & 20.1  & 6282.7  & -83.7 &  1.4\%  \\
       &  $6p^2 8p$    & 3/2&   6272.9  & 6414.7  & 6326.9  & 87.8  & 6.4  & 2.6  & 17.5  & 6353.7  & -80.8 &  1.3\%
\end{tabular}
\end{ruledtabular}

\flushleft $^1$ A. Kramida, Yu. Ralchenko, J. Reader, and the NIST ASD Team (2019). NIST Atomic Spectra Database (version 5.7.1). Available at http://physics.nist.gov/asd. National Institute of Standards and Technology, Gaithersburg, MD.

\end{table*}

 \begin{table*} [b]
\caption{Supplemental Material. Calculated energy levels of Po relative to the Po $^3\textit{P}_2$ ground state in meV.  The experimental values are from NIST$^1$.  Column definitions are the same as for Table~\ref{table1}.}
\begin{ruledtabular}
\begin{tabular}{lccccccc}
\multicolumn{1}{c}{Level}& \multicolumn{1}{c}{Expt.}& \multicolumn{1}{c}{CI+MBPT}&\multicolumn{1}{c}{CI+All}&  \multicolumn{1}{c}{HO}&
\multicolumn{1}{c}{Final}& \multicolumn{1}{c}{Diff.}&  \multicolumn{1}{c}{Diff.\%} \\
\hline
$6p^4$ ~$^3\textit{P}_2$  &       0  &     0   &     0  &     0 &     0   &   -     \\
$6p^4$ ~$^3\textit{P}_0$  &    931.7  &  994.6  &  959.5  &  -35.1  &  956.8  & -25.1 & 2.5\% \\
$6p^4$ ~$^3\textit{P}_1$  &    2086.9  &  2046.7  &  2007.2  &  -39.4  &  2017.5  & 69.4 & 3.4\%    \\
$6p^4$ ~$^1\textit{D}_2$  &    2687.9  &  2649.5  &  2626.4  &  -23.2  &  2626.1  & 61.8 & 2.3\%      \\
\end{tabular}
\end{ruledtabular}

\flushleft $^1$ A. Kramida, Yu. Ralchenko, J. Reader, and the NIST ASD Team (2019). NIST Atomic Spectra Database (version 5.7.1). Available at http://physics.nist.gov/asd. National Institute of Standards and Technology, Gaithersburg, MD.

\end{table*}

\end{document}